# Experimental Studies on Spatial Resolution of a Delay-Line Current-Biased Kinetic-Inductance Detector

The Dang Vu[1,2], Hiroaki Shishido[3], Kazuya Aizawa[2], Takayuki Oku[2], Kenichi Oikawa[2], Masahide Harada[2], Kenji M. Kojima[4], Shigeyuki Miyajima[5], Kazuhiko Soyama[2], Tomio Koyama[1], Mutsuo Hidaka[6], Soh Y. Suzuki[7], Manobu M. Tanaka[8], Masahiko Machida[9], Shuichi Kawamata[1] and Takekazu Ishida[1]

[1]Division of Quantum and Radiation Engineering, Osaka Metropolitan University, Sakai, Osaka 599-8570, Japan
[2]Materials and Life Science Division, J-PARC Center, JAEA, Tokai, Ibaraki 319-1195, Japan
[3]Equipment Sharing Center for Advanced Research and Innovation, Osaka Metropolitan University, Sakai, Osaka 599-8570, Japan
[4]Centre for Molecular and Materials Science, TRIUMF, 4004 Wesbrook Mall, Vancouver, BC V6T 2A3, Canada
[5]Advanced ICT Research Institute, National Institute of Information and Communications Technology, 588-2 Iwaoka, Nishi-ku, Kobe, Hyogo 651-2492, Japan
[6]Advanced Industrial Science and Technology, Tsukuba, Ibaraki 305-8568, Japan
[7]Computing Research Center, Applied Research Laboratory, High Energy Accelerator Research Organization (KEK), Tsukuba, Ibaraki 305-0801, Japan.
[8]Institute of Particle and Nuclear Studies, High Energy Accelerator Research Organization (KEK), Tsukuba, Ibaraki 305-0801, Japan.
[9]Center for Computational Science & e-Systems, JAEA, Kashiwa, Chiba 277-0871, Japan

ABSTRACT
A current-biased kinetic inductance detector (CB-KID) is a novel superconducting detector to construct a neutron transmission imaging system. The characteristics of a superconducting neutron detector have been systematically studied to improve spatial resolution of our CB-KID neutron detector. In this study, we investigated the distribution of spatial resolutions under different operating conditions and examined the homogeneity of spatial resolutions in the detector in detail. We used a commercial standard Gd Siemens-star pattern as a conventional method to estimate the spatial resolution and a lab-made $^{10}$B-dot array intended to examine detailed profiles on a distribution of spatial resolutions. We found that discrepancy in propagation velocities in the detector affected the uniformity of the spatial resolutions in neutron imaging. We analyzed the ellipsoidal line profiles along the circumferences of several different test circles in the Siemens-star image to find a distribution of spatial resolutions. Note that we succeeded in controlling the detector temperature precisely enough to realize stable propagation velocities of the signals in the detector to achieve the best spatial resolution with a delay-line CB-KID technique.

## 1. Introduction

Neutrons directly interact to nuclei [1] in contrast with the fact that the X-rays interact with electron cloud in atoms. Generally speaking, the absorption of X-rays has a monotonic dependence on the atomic number while neutrons are good at observing the light atomic-number elements such as hydrogen, lithium, boron and particular elements of high neutron absorption coefficients such as $^{10}$B, $^{113}$Cd, Gd. Neutron imaging technique has intensively been developed as a versatile non-destructive and non-invasive technique [2], and becomes a powerful tool to clarify the characteristics for a wide range of interesting materials in combination with the X-ray imaging [3]. Neutrons have been used for taking transmission spectra [2,4], for conducting three-dimensional imaging with neutron tomography [5], and for investigating pore structures in an electron beam-melted Ti-6Al-4V cube as were observed in the attenuation and dark-field images [6]. Therefore, neutron-imaging technology has been developed rather rapidly to realize high-resolution imaging, an energy-dispersive imaging, and a high-speed readout system of the signals from detectors. The best spatial resolution of 2 μm was reported in a scintillator-camera detector by using a CMOS sensor with the aid of center-of-gravity corrections for event coordinates [7]. The second-best spatial resolution was achieved by color-center formation in LiF crystals [8] as 5.4 μm. However, they are not perfectly suitable for the instrumentation under high-intensity pulsed neutrons. One of the suitable detectors is a $^{10}$B-doped microchannel plate (MCP) [9] so that it gives not only a high spatial resolution of 15 μm at low count rates but also a moderate resolution of 55 μm even at high count rates [10].

Our group proposed a current-biased kinetic inductance detector (CB-KID) [11], of which the operating principle is based on a rapid reduction in the local Cooper pair density $n_S$ at a tiny hot spot in the nanowire stripline with a very short length $\Delta \ell$ ($\ll$ total length of the meanderline $\ell$) when a neutron reacts with a $^{10}$B nucleus in the conversion layer. As a consequence of systematic studies on the CB-KID characteristics [12,13], we optimized the structure of superconducting neutron detector and the operating conditions to improve the spatial resolution from 22 μm [14] down to 16 μm [15]. However, the details on a distribution of spatial resolutions over the sensitive area have not been characterized well yet while the homogeneity of spatial linearity was investigated rather carefully as reported in our earlier publication [16]. Spatial resolutions of the neutron detectors have mostly been investigated using commercial fabricated test patterns such as a Gd-based Siemens star [17] and a 1951 USAF resolution test target with fine chrome bars [18]. Several lab-made test patterns have also been utilized by examining the sharpness of image edges of the test pattern with respect to background matrix area such as Cd patterns [19] and a $^{10}$B-dot array [14]. Commercial conventional patterns are useful to provide good comparisons between different detectors used at various research institutes, but it is not easy to understand distributions (or homogeneity) of the spatial resolutions over the sensitive area.

In this study, we experimentally examined the spatial resolution of CB-KID by using not only a commercial Gd Siemen star pattern but also a lab-made $^{10}$B-dots array to clarify the spatial resolution over the wide sensitive area of the CB-KID detector. The Gd Siemens star is suitable to compare objectively with reports on other detectors. We consider that a lab-made test sample is useful to clarify the position-dependent nature of the CB-KID sensor in view of spatial resolution and homogeneity. This assists us to reach a deeper understanding of the operating principle of the CB-KID sensor.

## 2. Experimental details

### 2.1. Superconducting Neutron Sensor

The CB-KID sensor is unconventional as a neutron detector [11] because it operates under an entirely different operating principle based on utilizing superconductive local dynamics compared to any existing neutron detectors. A device structure of CB-KID is very similar to that of a superconducting nanowire single photon detector (SNSPD) [20], but a bias current of CB-KID can be stayed at a level much smaller than the critical current while SNSPD requires a larger bias current close to the critical current. A microwave kinetic inductance detector MKID [21] utilizes a kinetic inductance change of the whole detector, but CB-KID utilized a local inductance change of the local hot spot. Our neutron detector consists of a superconducting ground plane, two orthogonal superconducting Nb meanderlines as the X and Y sensors [22]. We illustrate the CB-KID sensor by a schematic diagram in **Fig. 1(b)** and optical photos in **Fig. 1(c)**. We also need a $^{10}$B conversion layer deposited on top of the X and Y meanderlines to be used as a neutron detector. When a neutron reacts with a $^{10}$B nucleus in the conversion layer, one of the charged particles emitted from nuclear reaction creates a rapid reduction in the local Cooper pair density $n_S$ at a tiny hot spot of the nanowire stripline (meanderline) with a very short length $\Delta\ell$ (≪ total length $\ell$). This inducts a sudden enhancement in local kinetic inductance as an origin of signal emergence via superconducting dynamics in a restricted region. Under a small DC bias current $I_b$, a pair of pulsed-voltage signals is produced at the tiny hot spot in the very long stripline (meanderline) of the detector. The nanowire meanderline works for the two signals not only as a low-loss impedance-matched waveguide but also as an indispensable part of a delay-line-stripline detector for signals. The pulsed-voltage signals propagate toward both end-electrodes with opposite polarities as electromagnetic-wave pulses [23,24]. The time stamps of pulsed-voltage signals are recorded by Kalliope-DC readout circuit [25] having a 1-ns time-to-digital converter (TDC) after amplifying by low noise amplifiers (**Fig. 1(a)**). Time stamps are used to identify the position of neutron reaction (a hot spot) in detector to construct neutron image. **Figure 1(c)** shows the SEM images of the detector. The superconducting neutron sensor was fabricated on Si substrate in the following stacking sequence; (1) a 300 nm thick Nb ground plane, (2) a 350 nm $SiO_2$ layer, (3) a 50 nm thick Y meanderline, (4) a 150 nm thick $SiO_2$ layer, (5) a 50 nm thick X meanderline, and (6) a 150 nm thick $SiO_2$ layer at a superconductive foundry named as the clean room for analog-digital superconductivity (CRAVITY). The $^{10}$B neutron conversion layer of 220 nm thickness was deposited using electron-beam evaporation under ultra-high vacuum. The X and Y meanderlines had 10,000 repetitions of stripline segments ($h$ = 15.1 mm in length) with a pitch ($p$ = 1.5 μm) and the stripline width of 0.9 μm, giving a wide sensitive area of 15 mm × 15 mm. The turnaround points of the Nb segments are rounded by keeping the same width so that signals can transmit smoothly over a long superconducting meandered waveguide ($\ell$ = 151 m).

### 2.2 Measurement system for CB-KID

Neutron beams are not ideally parallel but have some divergence from the flight-path direction in the beam line. To avoid blurring of the neutron transmission image, we placed the test imaging samples at the very vicinity of the CB-KID sensor on the upstream side. This means that the samples were placed in the cryogenic environment at 0.8 mm (or at 0.7 mm in the case of the Gd Siemens star, see below) from the CB-KID meanderlines. This includes a 0.1-mm thick Al plate to mount imaging samples. Under beam conditions at the collimation ratio ($L/D$ = 140 or 335) of a rotary collimator at BL10 of Materials and Life Science Experimental Facility (MLF) in J-PARC over a flight path $L$ [26] and a collimator size of $D$, the best possible spatial resolution of the transmission image of the test sample at a cryogenic temperature is supposed to be 5.7 μm (5.0 μm in use of the Gd Siemens star) and 2.4 μm (2.1 μm in use of the Gd Siemens star) [27], respectively.

**Figure 1 (a)** shows a schematic diagram of the CB-KID measurement system. Both neutron sensor and test samples were installed on an anti-vibration cold stage of a Gifford–McMahon (GM) refrigerator to cool down to the operating temperature of 7.9K. The temperature of the superconducting sensor was controlled and monitored with a LabVIEW program by using two control loops of a controller (Cryocon Model 54), two electric resistance heaters, and two Cernox thermometers (Lake Shore Inc.) installed at the position near the neutron sensor and at the cold stage of the GM refrigerator. We used two variable-voltage sources to feed stable direct-current (DC) bias currents to the X and Y meanderlines through two bias resistors in series connection. Four voltage signals emerged from a single neutron event on the meanderlines were amplified using ultra-low-noise differential amplifiers (NF SA-430F5), and four 50-Ω signal splitters were used to divide the signals to input to two measurement instruments, i.e., a four-channel 2.5-GHz sampling digital oscilloscope (Teledyne LeCroy HDO4104-MS) and a 32-channel 1-GHz time-to-digital converter (TDC) module of the Kalliope-DC readout circuit [25]. The high-speed digital oscilloscope was used to tune/monitor the amplitudes of event signals by adjusting the magnitude of the bias currents feeding through the X and Y meanderlines while the Kalliope-DC readout circuit with temporal resolution of 1 ns is necessary to construct a neutron image by recording timestamps of the four signals with the aid of a delay-line technique. We used the 5-bit variable attenuators ATT (Hoshin Electronics N032) to adjust the signal amplitudes into the four TDC inputs channels of the Kalliope-DC readout circuit. The alternating-current (AC) powers of 100 V to the above instruments were provided from two stabilized AC power synthesizers (NF EC1000SA), i.e., we avoided notable environmental noises provided through the commercial AC power line at the large-scale accelerator facility. To reduce noises during transmittance of a high-frequency signal pulse from CB-KID in the cryostat at the cryogenic temperature to the readout instruments at room temperature, we used semirigid cables (50 Ω) with microminiature coaxial (MMCX) connectors and subminiature version A (SMA) connectors in the cryostat.

### 2.3. Samples to evaluate spatial resolutions of CB-KID

**Figure 2** shows the test samples used in this study to estimate the spatial resolution of superconducting neutron detectors. **Fig. 2(a)** is a photograph of a commercial Gd Siemens star (Paul Scherrer Institute (PSI)) with the Gd film of 5-μm thickness, which yields to absorb 63% of neutrons with wavelength of 2 Å [17] enough to observe neutron transmission image at high contrast. It has 128 Gd spokes radially extending outward from the center. The minimum distance between two adjacent spokes is 9.1 μm at the circumference edge of the central Gd island (dot) of radius 185 μm. The spatial resolution ($d$) using the neutron image of the Gd Siemens star was estimated by finding the minimum radius ($r$) of a test circle, on which we can resolve a separation ($d = 2\pi r/128$) of two adjacent Gd spokes. The advantage of employing the Gd Siemens star is that we can estimate the best spatial resolution of the detector to compare convincingly with other reports. However, it does not provide the detailed profile of the spatial resolutions along the x and y directions inside the detector although the detector has a wide sensitive area. It is meaningful to explore further details of spatial resolutions in our detector,

Alternatively, we prepared a 10×60 $^{10}$B-dot array by filling up very fine $^{10}$B particles into regular-hole array in a 50 μm-thick stainless-steel plate. Each hole has a diameter of 120 μm and a pitch of 250 μm of the $^{10}$B-dot array. The edge of the holes in the stainless-steel plate was somewhat rounded because holes were prepared by a wet etching process. We set a row of the 60 $^{10}$B dots at the position near the border of the sensitive area of the neutron detector along the 15 mm side of the $y$ direction to study the distribution of spatial resolutions as a function of the $y$ coordinate while a column of the 10 $^{10}$B dots at the left-edge side of detector to examine a distribution in spatial resolutions and to compare a difference of the spatial resolutions either along $x$ direction or along the $y$ direction. The Gd Siemens star was placed at a 700-μm distance from the CB-KID detector because it was 800 μm when it is necessary to use a 0.1-mm thick Al plate to hold measurement samples at a certain position. A spatial resolution of CB-KID sensor was the sum contribution not only from edge rounding of the stainless-steel hole containing fine $^{10}$B powder but also from blurring of the neutron image of the $^{10}$B dots.

Neutron irradiation was performed for 42 h at an operating temperature $T$= 7.9 K using a rotary collimator of $L/D$ = 140 (Large) and for 34 h at an operating temperature $T$ = 4.0 K using a collimator of $L/D$ = 335 (Medium) under a proton power of 640 kW for imaging the Gd Siemens star. We conducted the transmission imaging for 34.5 h at $T$=7.9 K with a collimator of $L/D$ = 335 (Medium) under a proton power of 915 kW for the $^{10}$B dots array at BL10 to obtain the transmission images.

We consider that an operating temperature at near critical temperatures (7.9 K) gives a smaller pixel $\Delta s_{x(y)} = \frac{p\, t_{res} v_{x(y)} N}{2L}$ because a slower velocity (see **Fig. 7(a)**) tends to give a superior spatial resolution. Here, $\Delta s_{x(y)}$, $v_{x(y)}$ are the pixel size, the propagation velocities of the voltage pulses propagating along the $x$ or $y$ direction; $t_{res}, p, N$ and $L$ are the temporal resolution, the pitch, the number segments of the meanderline and the length of the meanderline, respectively. However, a temperature controller at high-stability becomes more difficult to become real. If the temperature dependence of the velocity becomes steeper, the velocity fluctuations become more remarkable even under the same temperature fluctuations (see **Fig. 7(a)**). In this study, we intended to consider a question of what temperature should be the best to do the imaging experiments or of how accurate should a temperature stability be satisfied at higher operating temperatures. Those questions should be examined to fulfill high resolutions and high homogeneity in spatial resolutions over the sensitive area of the detector.

We note that this beam-power operation was almost as strong as the official specification of the J-PARC facility. In that sense, the present studies are indispensable for the development of the J-PARC facility.

## 3. Results and discussion

### 3.1. Spatial resolution of neutron image using the Gd Siemens star

**Figures 3(a)** and 3(**b**) are the neutron images of the Gd Siemens star obtained with CB-KID at different operating temperatures at 7.9 K and at 4.0 K, respectively. The image intensity composed of neutron energies from 230 μeV to 53 meV was normalized by the image intensity composed of higher neutron energies (> 53 meV) having smaller neutron absorption cross section. This method plays a similar role to the so-called open beam normalization, where the neutron image of the sample is normalized by using the image obtained under the identical conditions except for not placing the sample. This method is convenient because it is hard to conduct the open beam normalization when we place a sample at a cryogenic temperature near the CB-KID sensor. We proved that this method works rather well to remove blurring in the neutron transmission images in our earlier work [16]. By the comparison with **Fig. 3(b),** we found that a few vertical skewness lines are still there in **Fig. 3(a)** in the regions from $x$~−2 mm to $x$~− 1 mm and from $x$~1 mm to $x$~2 mm. We consider that skewness lines arise from the defects in meanderlines, which may be introduced in fabrication process of the sensor. The defect may work as a scatter for a signal during propagating waves inside the detector, and may prevent a voltage signal to propagate smoothly from the hot spot to the end electrode. This effect becomes more visible when a detector works at higher operating temperatures, and becomes less visible when a detector works at lower operating temperatures. This behavior appears at any wavelengths of neutrons, and hence it can partially be removed by the normalization of the image constructed with higher energies (or with lower absorption energies). We convincingly recognize that an image contrast between the Gd area and Gd-free area becomes more remarkable. The vertical line at $x = -1675$ μm appears due to the presence of defect. If the defect is not so serious to prevent signal transmission from passing through the defect in the CB-KID sensor, it may be from a weak link appearing in the meanderline. The weak link coming from a short thinner part in the meanderline may have a deteriorated critical current locally. We consider signal transmission would be influenced such as signal reflection or signal attenuation at the defect. This gives rise to blurring or skewness in image due to incorrect position assignment of the shot spot. A fatal defect is a disconnection of the meanderline so that any signal transmission across the defect is no longer possible.

To examine the best spatial resolution of our superconducting neutron detector, we attempted to find the smallest test circle in the neutron image of the Gd Siemens star **(Fig. 4(a))**, on which a periodic line profile of the Gd-spokes can be identified by the Fourier analyses. We took an ellipse-line profile of the image intensity along a circumference for several different circles with different radii in the inside of the first Gd ring, i.e., the first circle (in green) of radius $R = 900$ μm, the second circle "1/2" (in pink) of radius $R = 450$ μm, the third circle "1/3" (in blue) of the radius $R = 300$ μm, and the fourth circle "1/4" (in brown) of radius $R = 230$ μm. We took an extra circle to find the smallest circle (in yellow) of radius $R = 250$ μm, on which a Gd spoke is discernible from surrounding spokes. Ellipsoidal line profiles were taken as a function of central angle θ from 0 to 360 degrees in a counter-clock-wise direction, and were analyzed by the Fast Fourier Transformation (FFT) to judge whether a fundamental harmonic peak appears or not as a function of inverted angle (in frequency domain). The word of "ellipsoidal" is rather literal although it must be a circle in real space. This is because the pixel sizes are different between the *x* direction and the *y* direction. This arises from different propagation velocities of the signals because we used a delay-line technique in evaluating (*x,y*) coordinates. The fundamental peak corresponds to real-space oscillation with a period of $\theta = 360/128 = 2.8$ degrees in a line profile. **Fig. 4(b)** shows the FFT spectra from the circular-line profile of the image of the Gd Siemen star taken at 7.9 K (see **Fig. 4(a)**). The fundamental peak is apparent to be discerned in the largest circle ($R=900$ μm). The height of fundamental peak becomes weaker and weaker when the radius of the test circle becomes smaller towards the centeral Gd island. The position of the fundamental peak was evaluated by the Gaussian fitting at $1/\theta = 0.3556$. This corresponds to the angle of two adjacent spokes as $\theta = 2.793$ degrees. In **Fig. 4(b)**, the fundamental peak is still distinguishable at $R = 250$ μm (in yellow). This corresponds to the spatial resolution of $d = 12.27$ μm. In case of $R = 230$ μm (in brown), we find that the fundamental peak becomes not so evident. We magnified the intensity scale by a factor of 2 for two spectra (colors in brown and yellow) for ease of viewing. We conclude that the analyses of **Fig. 4** show the spatial resolution *d* of our CB-KID detector is 12.27 μm on the circumference of $R = 250$ μm.

### 3.2. Spatial resolution from neutron transmission image of the $^{10}$B dots array

**Figure 5(a)** shows the neutron transmission image of the 10×60 $^{10}$B-dots array. We found that the $^{10}$B-dots array was fabricated well in our laboratory enough to give a good contrast in the neutron transmission image between stainless-steel areas and the $^{10}$B dots. **Figure 5(b)** is the zoomed image of the $^{10}$B dots array while **Fig. 5(c)** is a line profile along the dashed line (in yellow) to evaluate the spatial resolution of neutron image from the sharpness of the edge images of the $^{10}$B dots. We found the interesting results that **Fig. 5(d)** and **Fig. 5(e)** were obtained by analyzing all $^{10}$B dots both along the *x* and *y* directions. The line profiles in **Fig. 5(c)** were fitted by an equation $I(x) = I_0 + A\left[\tanh\left(\frac{x-x_1}{x_w}\right) - \tanh\left(\frac{x-x_1-x_s}{x_w}\right)\right]$, where $I_0$ is the floor intensity, $A$ is an amplitude of the peak (or trough), $x_1$ is the position of the peak (or trough), $x_w$ is the width of the $^{10}$B dot edge, and $x_s$ is a measure of the size of the $^{10}$B dot. This function is convenient for modeling the various shapes of the line profiles across the dot, because it can fit both narrow and wide peaks (troughs), and peaks (troughs) with sharp or flat summits (minima). The distribution of spatial resolution of neutron image along the *y* direction is showed in **Fig. 5(d)**, where the spatial resolutions were calculated by full width at half maximum FWHM of the edge width as $\text{FWHM} = \ln(3 + 2\sqrt{2})x_w$, and each color and symbol style represent to a vertical line profile across the $^{10}$B dots.

In **Fig. 5(d)**, the best spatial resolution of the CB-KID sensor appears at the center of the detector, and the resolution monotonically decreases as a position moves toward both edges of the detector at $x = \pm 7.5$ mm. We found that a sequential change of a spatial resolution distribution as a function of position from center to edges of detector. This must come from a slight discrepancy of the velocity in evaluating the hot-spot positions. The positions of hot spot along the *x* and *y* directions are given by $x = \frac{t_{x-} - t_{x+}}{2h} v_x p$ and $y = \frac{t_{y-} - t_{y+}}{2h} v_y p$, where *x*, *y*, $v_x$, $v_y$ are the positions of hot spot and propagation velocities of the voltage pulses propagating along the *x* and *y* directions, respectively; $t_{x-}, t_{x+}, t_{y-},$ and $t_{y+}$ are the signal arrival timestamps; *p, h* are the pitch and the segment of the meanderline. We investigated the *Y* meanderline by a relation $y + \Delta y = \frac{t_{y-} - t_{y+}}{2h}(v_y + \Delta v_y)p$; where $\Delta y$ is the error in position of hot spot arising and $\Delta v_y$ is the error in propagation velocity. The relation $\Delta y = \frac{t_{y-} - t_{y+}}{2h}\Delta v_y p$ explains why the spatial resolution was deteriorated as a position approaches toward the detector edge. This is because a difference between $t_{y-}$ and $t_{y+}$ timestamps becomes maximum at the detector edge, but it becomes zero at the detector center. Considering an error of spatial resolution as $\Delta y = \frac{\Delta v_y}{v_y} y$, we obtain the relation $d + \Delta y = d + \frac{\Delta v_y}{v_y} y$. We fitted all the data points of spatial resolutions from the 10 curves in **Fig. 5(d)** by a linear relation $D = a \times |y| + d$ to estimate $d = 8.92$ μm and $\Delta v_y / v_y = 1.8 \times 10^{-3}$. Note that the neutron image of $^{10}$B dots was obtained with *L/D* = 335 at rotary-collimator setting of Middle position, which gives an improved spatial resolution compared to 12.27 μm of the Gd Siemens star taken at 7.9 K under *L/D* = 140 at rotary-collimator setting of "Large" position.

The spatial resolution of CB-KID evaluated from the neutron image with the $^{10}$B-dots array was in good agreement with that evaluated from the Gd Siemens star. The analysis method using $^{10}$B-dots matrix tended to indicate systematic deviation arising from an error in the propagation velocity used in evaluating spatial resolutions. We argue that it is necessary to minimize the error in estimating the propagation velocity. This comes partly from the velocity measurement and partly from the velocity fluctuations during the imaging data acquisition arising from the temperature fluctuations of the detector [13]. The CB-KID method requires further improvements in temperature stability to achieve a better spatial resolution.

**Fig. 5(e)** shows a variation of the spatial resolution along the $x$ direction. However, these $^{10}$B dots did not distribute along the $x$ direction wide enough to discuss a systematic tendency in the errors in the propagation velocity compared to the $x$ direction. We argue that the results along the $y$ direction would basically be identical to those along the $x$ direction because the CB-KID detector is completely symmetric with respect to the $x$ and $y$ directions in design. We obtain a spatial resolution of 19.02±1.85 μm along the $y$ direction by averaging all the data points in **Fig. 5(e)**.

### 3.3. Distribution of spatial resolutions using the Gd Siemens star

An error in propagation velocity comes not only from temperature stability in taking neutron image but also from a choice of an operating temperature. The Maxwell-London theory gave the equation between propagation velocity $v$ and operating temperature $T$ by $v = \frac{c}{\sqrt{\epsilon}}\sqrt{\frac{d}{d+\lambda_L(1+\coth(s/\lambda_L))}}$ , where $c$ is the speed of light in a vacuum, $\epsilon$ is a dielectric constant of SiO$_2$ insulator, $d$ is a thickness of SiO$_2$ insulating layer, $s$ is a thickness of the nanowire, respectively [23]. The London penetration depth $\lambda_L$ is expressed by the two-fluid model as $\lambda_L(T) = \lambda_L(0)/\sqrt{1-(T/T_c)^4}$ , where $T_c$ is a critical temperature and $T$ is a temperature of superconducting nanowire. This relation gives us a guide to understand why the stability of propagation velocity in CB-KID becomes better at lower operating temperatures because the temperature dependence of the velocity becomes weaker as $T$ decreases.

This is meaningful to evaluate how the error in the propagation velocity influences to diminish a spatial resolution of neutron image obtained by analyzing the sharpness of the neutron image at the edges. **Figure 3(b)** shows the neutron image of the Gd Siemens star obtained at 4 K. An advantage of employing neutron images observed at lower temperatures is due to the smaller effect of defects in the detector on the transmission image. Note that it is not possible to use the CB-KID detector having a fatal defect so that it is not possible to feed a bias current across a disconnected defect in a superconducting stripline. The defect profiles might be from superconducting weak links along the meanderline, but a superconducting coupling across the weak links becomes stronger as $T$ decreases. This explains the phenomena that the defect becomes less visible in transmission images at lower temperatures.

We used a hyperbolic tangent function to fit an ellipse-line profile of the neutron transmission image of the Gd spokes. This is an alternative approach in addition to the FFT analyses (see above). We skipped to take data points from the defect area at around $x = -1.67$ mm appearing between the first Gd ring and the second Gd ring to avoid the defect influence in evaluation. In **Fig. 6**, a FWHM width $w$ in a differential-curve peak of the edge slope at the Gd boundary in the Gd spokes was obtained as a function of test-circle radius $R$. Neutron images measured at 4 K give a weaker dependence as a function of radius $R$ by $w = 1.5 \times 10^{-4}\ R + 11.08$ (μm). An FWHM width $w$ at 7.9 K is strongly dependent on $R$ (as expressed by $w = 7.8 \times 10^{-4}\ R + 12.4$ (μm)) by a factor of 5.2. Regarding comparisons of the FFT analysis and the hyperbolic-tangent-function fitting, the best resolution of image 11.08 μm at 4 K was in reasonable agreement with 12.4 μm at 7.9 K. This is from a difference in neutron-beam parallelism with $L/D$ = 335 [27].

The influence of temperature on spatial resolution was checked by measuring the signal velocity as a function of $T$ from 4.0 K to 8.3 K (see **Fig. 7(a)**). The experimental points were fitted by a $v$ equation as a function of $T$ [23]. **Figures 7(b)** and **7(c)** are the histograms of temperature stability during imaging at 4 K and at 7.9 K. We fitted the histograms by Gaussian function to obtain the temperature stability as $T$ = 4 K ± 4.34 mK and $T$ = 7.9 K ± 0.62 mK, respectively. The velocity fluctuations are evaluated as $\Delta v_{x\_4K} = 1.191165 \times 10^4$ (m/s) and $\Delta v_{y\_4K} = 1.28014 \times 10^4$ (m/s) at 4 K and $\Delta v_{x\_7.9K} = 2.7273107 \times 10^4$ (m/s) and $\Delta v_{y\_7.9K} = 3.39615 \times 10^4$ (m/s) at 7.9 K. We evaluated the ratios as $\Delta v_{x\_4K}/v_{x\_4K} = 1.4 \times 10^{-4}$ , $\Delta v_{y\_4K}/v_{y\_4K} = 1.9 \times 10^{-4}$; $\Delta v_{x\_7.9K}/v_{x\_7.9K} = 4.8 \times 10^{-4}$ and $\Delta v_{y\_7.9K}/v_{y\_7.9K} = 9.3 \times 10^{-4}$. We obtained $\Delta v_{4K}/v_{4K} = 1.65 \times 10^{-4}$ and $\Delta v_{7.9K}/v_{7.9K} = 7.05 \times 10^{-4}$ by averaging the results for the X and Y directions. These results are in good agreement with the results ($\Delta v_{4K}/v_{4K} = 1.5 \times 10^{-4}$ at 4 K and $\Delta v_{7.9K}/v_{7.9K} = 7.05 \times 10^{-4}$ at 7.9 K) (see **Fig. 6**). We conclude that the high stability in temperature control is very essential to extract the best performance of the CB-KID measurement and homogeneity in spatial resolutions in the detector.

## 4. Summary

In this work, we prepared two different types of test samples, i.e., one is a commercial Gd-based Siemens star and the other is a lab-made $^{10}$B-dots array to examine a systematic change in spatial resolutions of the CB-KID sensor. The spatial resolution of CB-KID was obtained by analyzing an ellipsoidal-line profile in the transmission image data by FFT. Note that this line profile is taken along a test-circle circumference in real coordinate. This method was much more convincing than the method of distinguishing each Gd spoke by the eye. An alternative method to estimate the spatial resolution of the CB-KID sensor was to use the neutron transmission image of the $^{10}$B-dots array by fitting the line profiles of $^{10}$B dots for finding the sharpness of the image at the edge between the $^{10}$B dots and the stainless-steel-plate holes. This method of using the $^{10}$B-dot array has an advantage because it provides detailed distribution in spatial resolutions over the sensitive area of the CB-KID sensor. The spatial resolution of neutron detector was confirmed to be better than 12 μm both using the Gd Siemens star and using the [10]B-dots array. A hyperbolic tangent function was used to analyze the circular-line profiles of the Gd Siemens star, and we obtain spatial resolution as a function of test circle $R$. We found that systematic involvements of the velocity fluctuations were relevant to the results in evaluating hot-spot positions of the ($x, y$) coordinates because the velocity fluctuations come from the temperature instability of the CB-KID detector. Therefore, further improvements in the temperature stability of the detector are essential to improve the system performance toward the best spatial resolution. We consider that a suitable temperature stability must provide the ratio $\Delta v/v$ small enough to observe the changing spatial resolution from center to edge of neutron image lower than a pixel size Since the CB-KID sensor can be operated over at an arbitrary temperature below a critical temperature $T_c$, it might be possible to find better operation conditions of the CB-KID sensor.

**Acknowledgements:** This work was partially supported by Grant-in-Aid for Scientific Research (A) (No. JP21H04666) and Grant-in-Aid for Early-Career Scientists (No. JP21K14566, JP23K13690) from JSPS. The experiments at the Materials and Life Science Experimental Facility (MLF) of the J- PARC were supported by the MLF project program (No. 2021P0501).

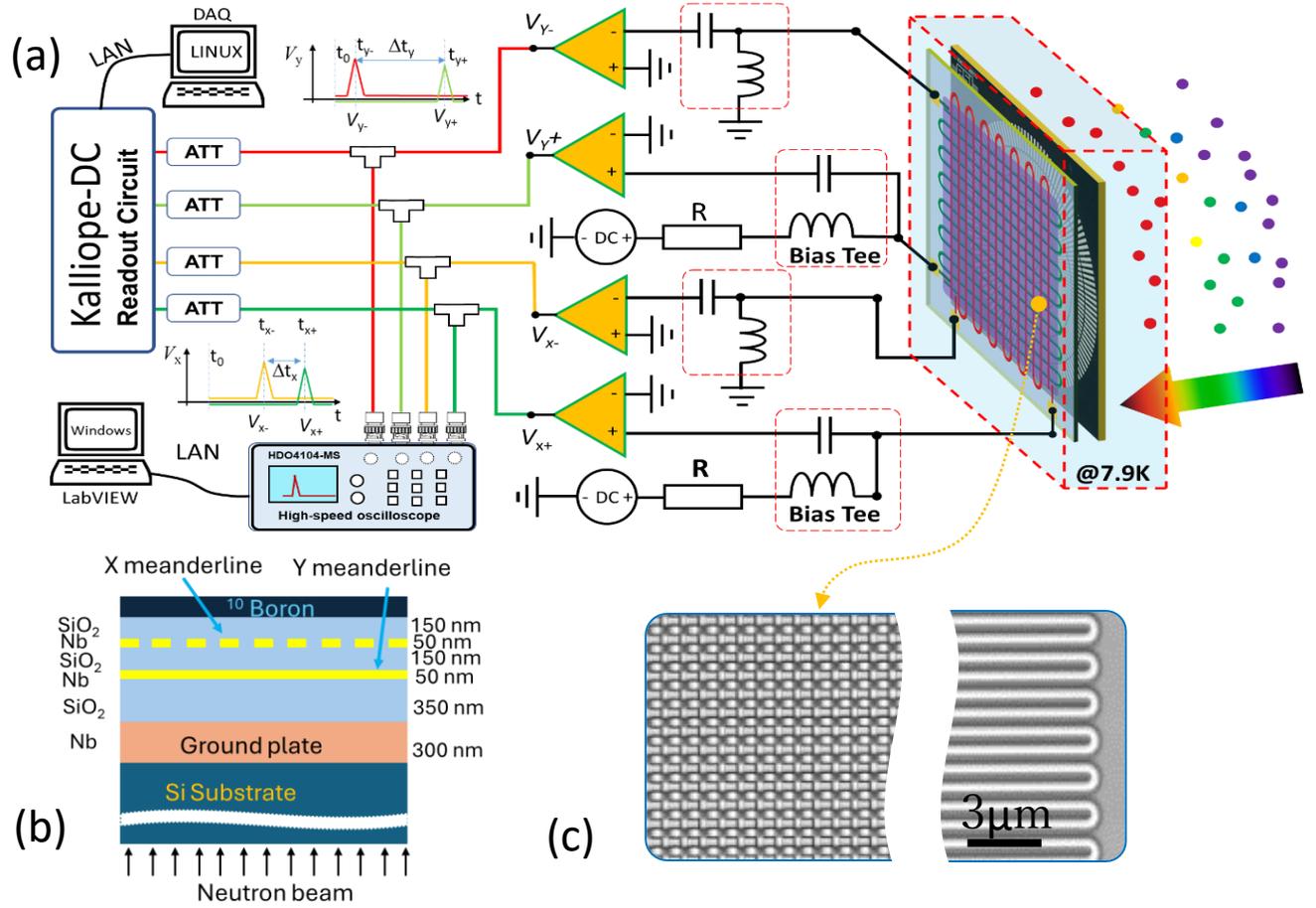

**Fig. 1(a).** Schematic diagram of the CB-KID measurement system. Firstly, pulsed neutrons at beam line BL10 of the MLF at J-PARC were transported along the 14-m long beamline. After passing through the test samples, a neutron incidents from the substrate side to the CB-KID sensor and reacts with a $^{10}$B nucleus in a $^{10}$B conversion layer to emit charged ions for producing hot-spot events in both X and Y meanderlines. A Gifford–McMahon cryocooler was used to cool down both the detector and the test samples to a cryogenic temperature of 7.9 K. The neutron detector consists of the X- and Y-meanderlines, which are superimposed orthogonally to each other, and a $^{10}$B neutron conversion layer. Neutron signals arising from both ends of the two meanderlines were amplified by ultra-low noise amplifiers at each channel (Ch1, Ch2, Ch3, or Ch4) to feed into a Kalliope-DC readout circuit and a digital oscilloscope via four signal splitters. The system was controlled by a data acquisition (DAQ) program and a LabVIEW software; **(b)** the schematic diagram of the cross-section of the X and Y CB-KID system; **(c)** SEM image shows the structure of X- and Y- meanderlines before covering by a $^{10}$B layer.

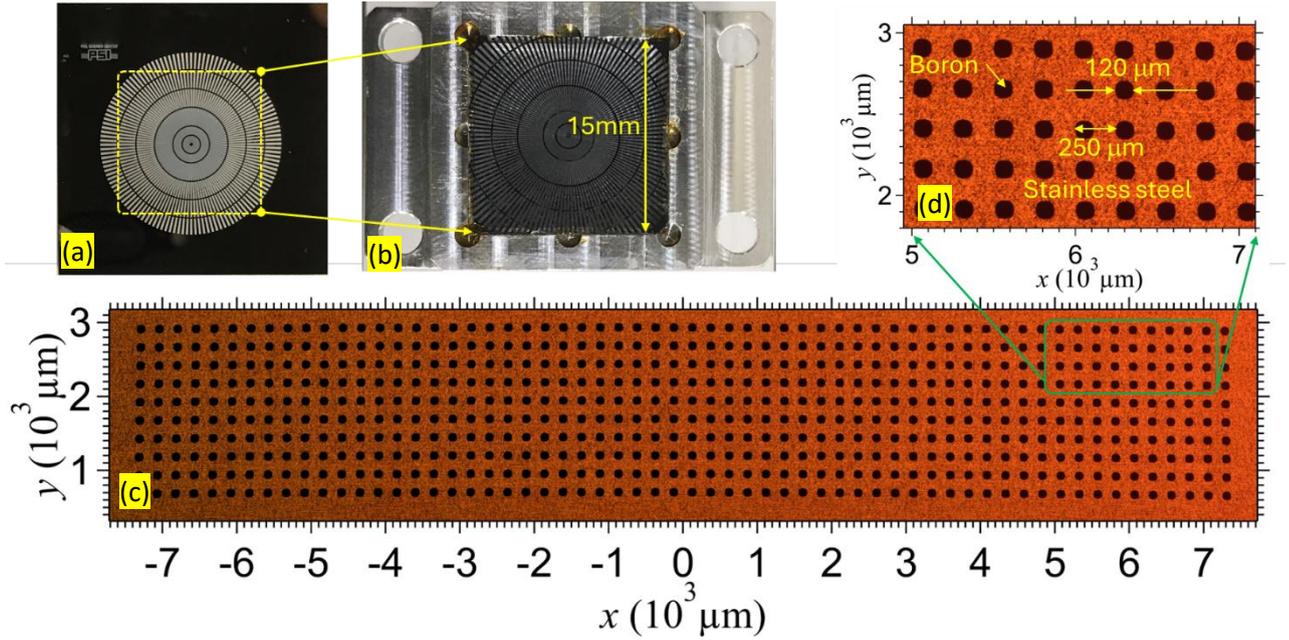

**Fig. 2.** Optical photographs of the test samples used in studying the spatial resolutions in neutron transmission image with the CB-KID system. **(a)** A commercial Gd Siemens star of 5-μm thickness with 128 Gd spokes. Fabrication of the Gd Siemens star is fine enough to estimate the best resolution down to 8.6 μm. **(b)** The Gd Siemens star was diced by a diamond saw to a size of the sample holder so at to fit the sensitive area 15 mm × 15 mm of the CB-KID sensor. **(c)** Optical photograph of the whole view of 10 × 60 $^{10}$B-dots arrays prepared by our laboratory with thickness of 50 μm and diameter of 120 μm. **(d)** Zoomed photograph of the $^{10}$B-dots array in the square in green in **Fig. 2(c).** The $^{10}$B-dots array has a period of 250 μm.

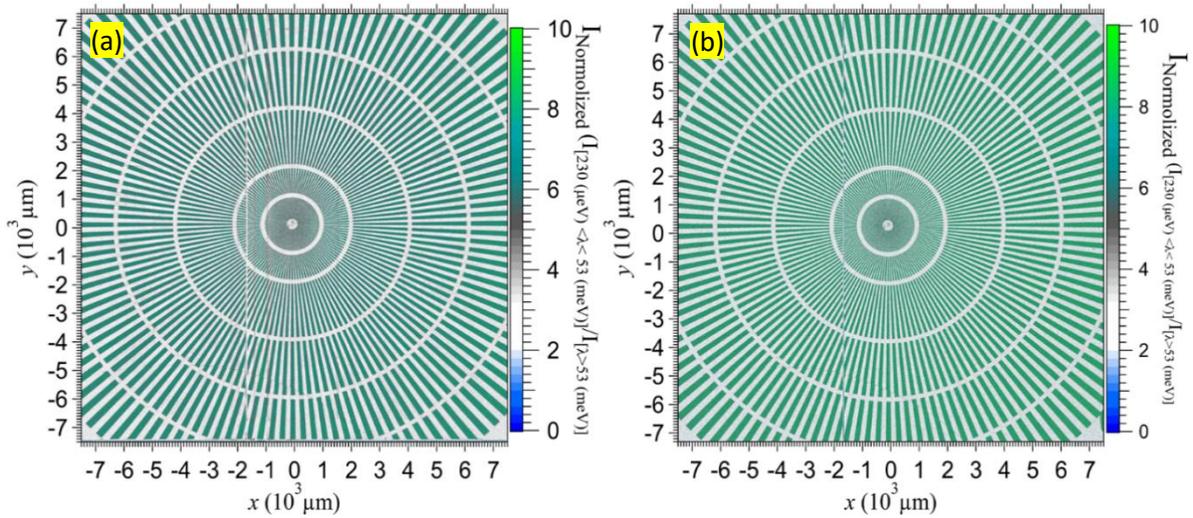

**Fig. 3.** Neutron transmission image of the Gd Siemens star **(a)** at 7.9 K and **(b)** at 4.0 K acquired with CB-KID. Neutron transmission images were composed with energies from 230 μeV to 53 meV, and were normalized by the image composed with neutrons of energies larger than 53 meV. This was useful to remove blurring/skewness in neutron transmission image because Gd has smaller absorption of neutrons below 53 meV. Some blurring is more visible when the image is taken **(a)** at 7.9 K (see text, too) than the image taken **(b)** at 4.0 K.

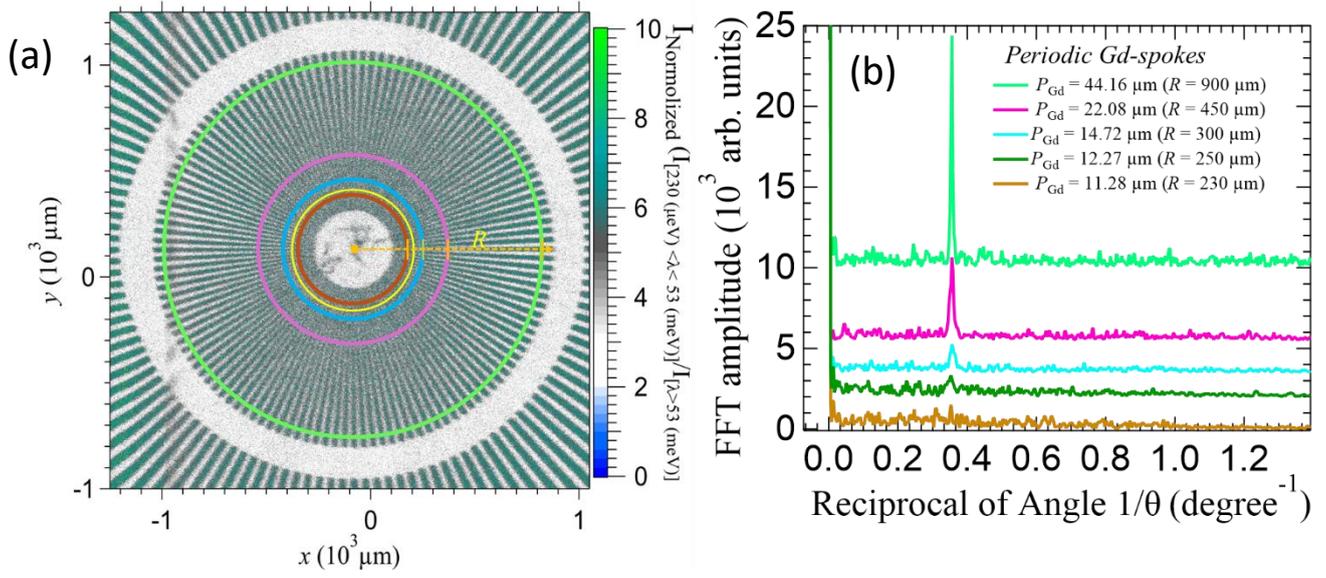

**Fig. 4. (a)** Magnified neutron transmission image of **Fig. 3(b)** at the central region of the Gd Siemens star and the several test circles to take the circular-line profiles as a function of angle θ. **(b)** The fast Fourier transformer (FFT) spectrum of the ring-line profiles around the test circles to judge whether the CB-KID resolves each Gd spoke or not. We consider the appearance of the fundamental peak in FFT spectrum in **(b)** is convincing evidence of discernibility of the Gd spokes in neutron transmission image.

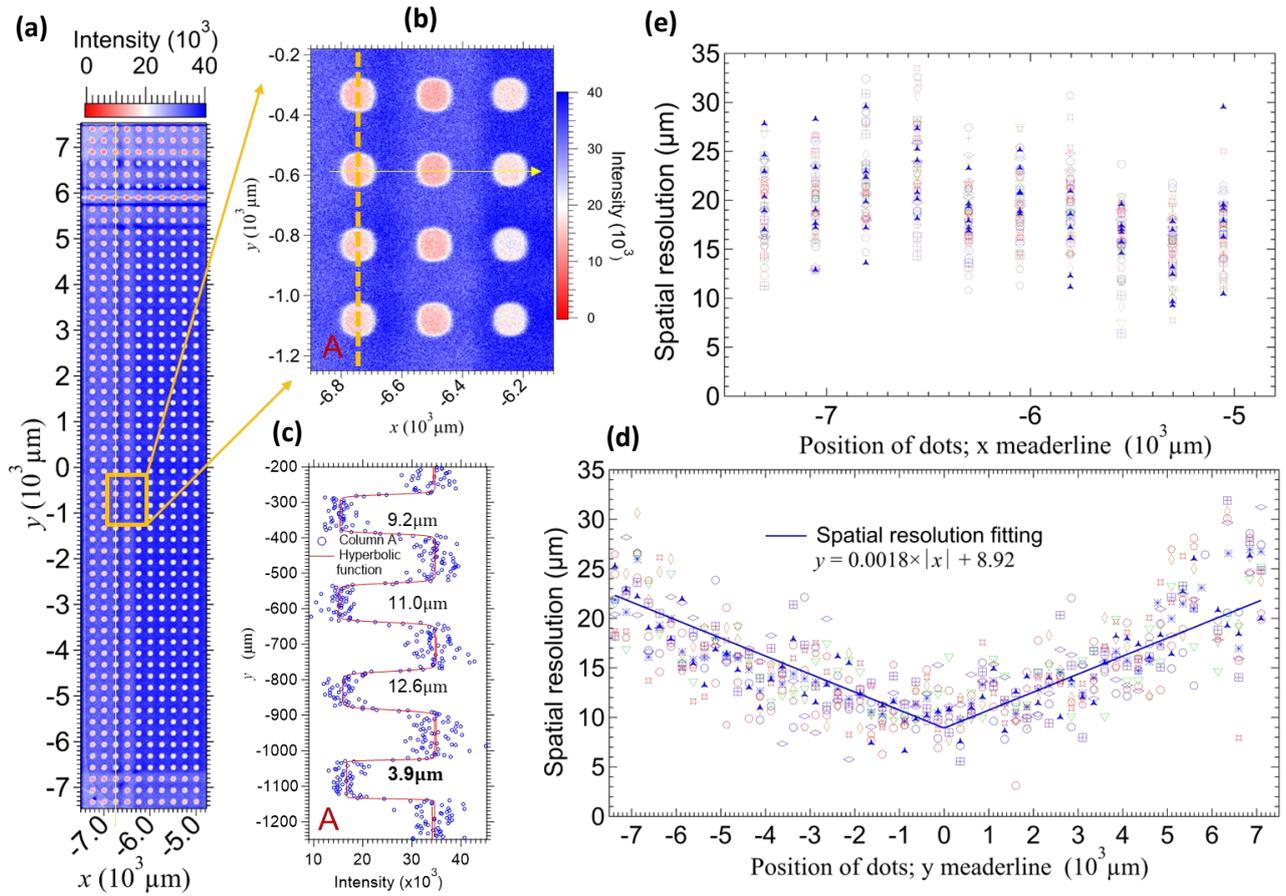

**Fig. 5.** The method of estimating spatial resolution of neutron image of the CB-KID sensor by means of the [10]B dots array. **(a)** Neutron transmission image of the [10]B dots array. **(b)** Magnified neutron image of the [10]B dots in a rectangular area (in orange) in **(a)**. **(c)** A vertical line profile of [10]B dots along the dashed line (in yellow) in **(b)** and a fitting curve by using a hyperbolic tangent function to evaluate the width of the edge slope between the [10]B dots and stainless-steel-plate holes. **(d)** A spatial resolution as a function of position along the $x$ direction. **(e)** A spatial resolution as a function of position along the $y$ direction.

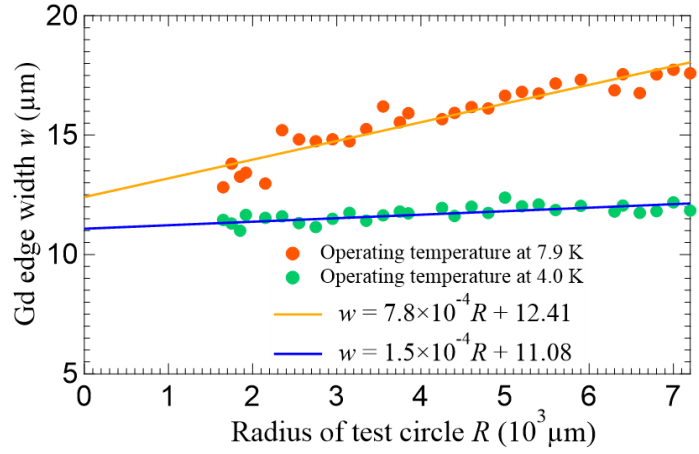

**Fig. 6.** The spatial resolution evaluated from the width of the Gd edge at 4.0 K and at 7.9K as a function of test-circle radius *R*. We defined a spatial resolution as a full width at half maximum (FWHM) in the differential peak curve of the fitting curve of the Gd edge given in a hyperbolic tangent function. A clear dependence of a spatial resolution as a function of *R* is due to different propagation-velocity fluctuations between 4.0 K and 7.9 K.

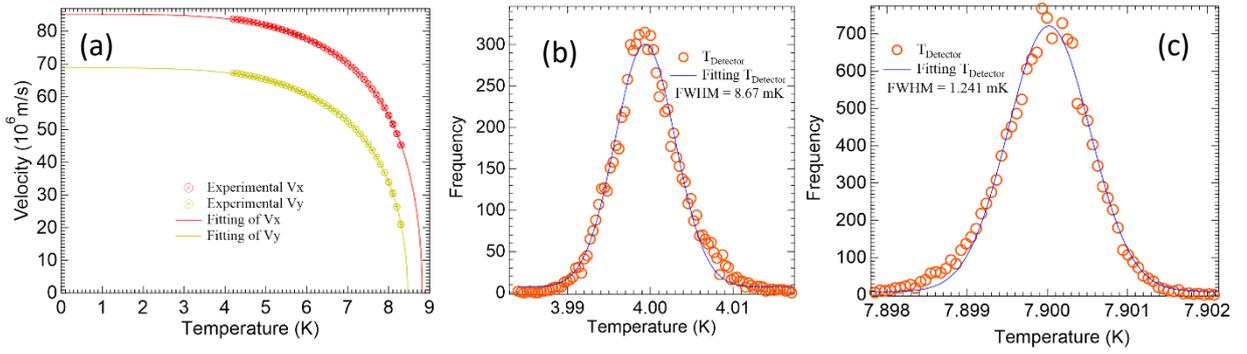

**Fig. 7. (a)** Propagation velocity measured as a function of temperature both along the *x* direction and along the *y* direction. The solid lines are the fitting curves by the theoretical formula of the two-fluid model [21]. **(b)** Temperature stability of CB-KID system at 4 K±4.34 mK. **(c)** Temperature stability of CB-KID system at 7.9 K±0.62 mK.